\documentclass[12pt]{article}
\usepackage{amsmath}
\usepackage{amsthm}
\usepackage{amsfonts}
\usepackage{algorithm}
\usepackage{algorithmic}
\usepackage[braket]{qcircuit}

\newtheorem{fact}{Fact}

\theoremstyle{definition}

\title{Quantum circuits for permutation matrices}
\author{jason hanson}

\begin{document}
\maketitle

\begin{abstract}
Two different algorithms are presented for generating a quantum circuit realization of a matrix representing a permutation on $2^n$ letters.  All circuits involve $n$ qubits and only use multi--controlled Toffoli gates.  The first algorithm constructs a circuit from any decomposition of the permutation into a product of transpositions, but uses one ancilla line.  The second, which uses no ancillae, constructs a circuit from a decomposition into a product of transpositions that have a Hamming distance of one.  We show that any permutation admits such a decomposition, and we give a strategy for reducing the number of transpositions involved.
\end{abstract}

\section{Introduction}

Given a non--negative integer $N$, define $[[N]]\doteq\{0,1,\dots,N-1\}$. An $n$--bit integer $I\in[[2^n]]$ will be represented in binary as $I=I_{n-1}\cdots I_1I_0$.  I.e., , $I=I_{n-1}\cdot 2^{n-1}+\cdots+I_1\cdot 2^1+I_0\cdot 2^0$.  The corresponding $n$--qubit computational basis element is $\ket{I}_n=\ket{I_{n-1}}\otimes\cdots\otimes\ket{I_1}\otimes\ket{I_0}$.

The symbol $\oplus$ will be used for bit--wise addition modulo two (XOR).  The Hamming distance $b(I,J)$ between two integers $I,J$ is defined to be the number of bits by which their binary representations differ.  Equivalently, the number of nonzero bits in $I\oplus J$.  For example, $3\oplus 6=011_2\oplus 110_2=101_2=5$, and $b(3,6)=2$.

The reader is assumed to be familiar with cycle notation for permutations.  And we will make use of the fact that every permutation is a non--unique product of transpositions.   For instance, two possible decomposition of the cycle $(0,2,1,3)$ are $(0,2)(2,1)(1,3)$ and $(0,3)(0,1)(0,2)$.

\subsection{Permutation matrices}
Let $\pi$ be a permutation of $[[N]]$.  We have an associated unitary $N\times N$ matrix $U_\pi$ whose elements are defined to be $(U_\pi)_{rc}=\delta_{r,\pi(c)}$.  That is, the entries of the $c$--th column of $U_\pi$ consist entirely of zeros, except for row $r=\pi(c)$, which has a value of unity.  E.g., for $N=4$, and $\pi=(0,2,1,3)$.  The associated unitary  matrix is
$$U_\pi=\left(\begin{smallmatrix} 0 & 0 & 0 & 1\\
                                  0 & 0 & 1 & 0\\
                                  1 & 0 & 0 & 0\\
                                  0 & 1 & 0 & 0
              \end{smallmatrix}\right).
$$

In quantum computation, a swap gate transposes the values of two qubit wires.  For $n$ qubits, a swap then is a transposition of $[[n]]$.  Decomposing into transpositions, we see that every permutation of the qubit wire values involves at most $n-1$ swap gates.

In contrast, our interest is in permutations of the computational basis itself, not just the qubit wire values.  I.e., permutations $\pi$ of $[[2^n]]$, which have the associated unitary matrix $U_\pi$ determined by
$$U_\pi\ket{I}_n=\ket{\pi(I)}_n,
  \quad\text{for $I\in[[2^n]]$.}
$$
The realization of such a permutation typically requires significantly more gates than a simple permutation of the qubit wire values.

\subsection{Motivation and related work}
Many basic quantum gates are permutation matrices, for instance the Pauli $X$ and Toffoli gates.  Moreover the oracle matrices used by the Deutsch--Jozsa and Simon quantum algorithms are permutations.  As are the underlying unitary operators used in Shor's and Grover's algorithms.

Shende et al. \cite{Shende} investigated quantum circuits without ancillae for some restricted classes of permutations, namely those that can constructed from NOT, CNOT, Toffoli, and swap gates only.  Adhikari \cite{Adhikari}, as part of their exploration of quantum circuits and random samplings of permutations, decomposes a permutation into numerically adjacent transpositions, those of the form $(j,j+1)$, in order to obtain a quantum circuit representation using no ancillae.

The methods presented in this article place no restrictions on the permutation type, but at the expense of using multi--controlled Toffoli gates (which can be decomposed into simpler gates, see \cite{Miller} for example).  The circuits of the first type, which use a single ancilla, are applicable to any decomposition of a permutation into transpositions.  On the other hand, the circuits of the second type, which use no ancillae, use bit--wise adjacent transpositions in the decomposition of a permutation.  Such transpositions are of the form $(I,J)$ where $I$ and $J$ have a Hamming distance of one, and can be represented by a single multi--controlled Toffoli gate.

\section{Permutation circuits with a single ancilla}\label{sec:oneancilla}

Using a single ancilla qubit, a possible representation of the transposition $(I,J)$ for $I,J\in[[2^n]]$ is given by the three stage reversible quantum circuit
\begin{enumerate}
\item\label{alg1:step1} $\ket{\psi}\otimes\ket{K}_n
       \mapsto\ket{\psi\oplus\delta_{KI}\oplus\delta_{KJ}}\otimes\ket{K}_n$
\item\label{alg1:step2} $\ket\psi\otimes\ket{K}_n
       \mapsto\ket\psi\otimes\ket{K\oplus(I\oplus J)^\psi}_n$
\item\label{alg1:step3} repeat stage \ref{alg1:step1}
\end{enumerate}
Where it is assumed that the ancilla in the initial state is $\ket\psi=\ket{0}$.  Evidently, stage \ref{alg1:step1} uses two multi-controlled Toffoli gates, and stage \ref{alg1:step2} uses $b(I,J)$ CNOT gates.  Stage \ref{alg1:step3} guarantees a clean ancilla ({\em strict clean non--wasting} in the classification in \cite{Khandelwal}), so that circuits can be chained to realize any decomposition of a permutation into transpositions.

As an example, the quantum circuit for the transposition $\tau=(5,6)$ on three qubits is\par
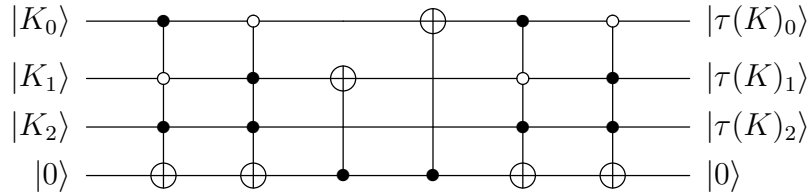
\centerline{\Qcircuit @R=12pt {
  \lstick{\ket{K_0}}
    & \ctrl{1} & \ctrlo{1} & \qw & \targ & \ctrl{1} & \ctrlo{1}
    & \rstick{\ket{\tau(K)_0}}\qw \\
  \lstick{\ket{K_1}} & \ctrlo{1}
    & \ctrl{1} & \targ & \qw & \ctrlo{1} & \ctrl{1}
    & \rstick{\ket{\tau(K)_1}}\qw \\
  \lstick{\ket{K_2}}
    & \ctrl{1} & \ctrl{1} & \qw & \qw & \ctrl{1} & \ctrl{1}
    & \rstick{\ket{\tau(K)_2}}\qw \\
  \lstick{\ket{0}}
    & \targ & \targ & \ctrl{-2} & \ctrl{-3} & \targ & \targ
    & \rstick{\ket{0}}\qw
}}
\medskip
\noindent
Note that $5=101_2$ and $6=110_2$, so that $5\oplus 6=011_2$.  Whence stage \ref{alg1:step2} is realized by $b(5,6)=2$ CNOT gates.

We remark that for cycles with length greater than 2, there is a circuit simplification that can be employed by canceling adjacent pairs of identical multi--controlled Toffoli gates.  For example, with the decomposition of the cycle $\sigma=(3,6,5)=(3,6)(6,5)=(6,3)(5,6)$, the circuit realization\par
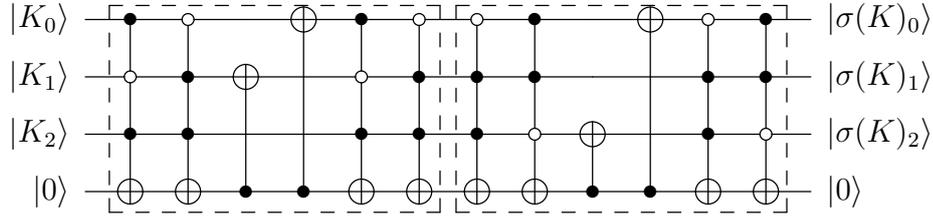
\centerline{\Qcircuit @R=12pt @C=12pt{
  \lstick{\ket{K_0}}
    & \ctrl{1}  & \ctrlo{1} & \qw  & \targ  & \ctrl{1}  & \ctrlo{1}
    & \ctrlo{1} & \ctrl{1}  & \qw  & \targ  & \ctrlo{1} & \ctrl{1}
    & \rstick{\ket{\sigma(K)_0}}\qw \\
  \lstick{\ket{K_1}}
    & \ctrlo{1} & \ctrl{1} & \targ & \qw & \ctrlo{1} & \ctrl{1}
    & \ctrl{1}  & \ctrl{1} & \qw   & \qw & \ctrl{1}  & \ctrl{1}
    & \rstick{\ket{\sigma(K)_1}}\qw \\
  \lstick{\ket{K_2}}
    & \ctrl{1} & \ctrl{1}  & \qw   & \qw & \ctrl{1} & \ctrl{1}
    & \ctrl{1} & \ctrlo{1} & \targ & \qw & \ctrl{1} & \ctrlo{1}
    & \rstick{\ket{\sigma(K)_2}}\qw \\
  \lstick{\ket{0}}
    & \targ & \targ & \ctrl{-2} & \ctrl{-3} & \targ & \targ
    & \targ & \targ & \ctrl{-1} & \ctrl{-3} & \targ & \targ
    & \rstick{\ket{0}}\qw
  \gategroup{1}{2}{4}{7}{6pt}{--}
  \gategroup{1}{8}{4}{13}{6pt}{--}
}}
\medskip
\noindent
(composition of cycles is right to left, while composition of circuits is left to right) contracts to\par
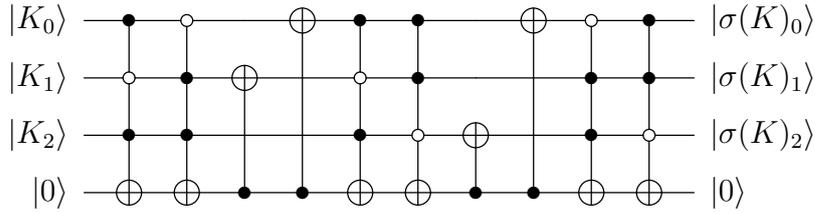
\centerline{\Qcircuit @R=12pt @C=12pt{
  \lstick{\ket{K_0}}
    & \ctrl{1}  & \ctrlo{1} & \qw   & \targ     & \ctrl{1}
    & \ctrl{1}  & \qw       & \targ & \ctrlo{1} & \ctrl{1}
    & \rstick{\ket{\sigma(K)_0}}\qw \\
  \lstick{\ket{K_1}}
    & \ctrlo{1} & \ctrl{1} & \targ & \qw      & \ctrlo{1}
    & \ctrl{1}  & \qw      & \qw   & \ctrl{1} & \ctrl{1}
    & \rstick{\ket{\sigma(K)_1}}\qw \\
  \lstick{\ket{K_2}}
    & \ctrl{1}  & \ctrl{1} & \qw & \qw      & \ctrl{1}
    & \ctrlo{1} & \targ    & \qw & \ctrl{1} & \ctrlo{1}
    & \rstick{\ket{\sigma(K)_2}}\qw \\
  \lstick{\ket{0}}
    & \targ & \targ     & \ctrl{-2} & \ctrl{-3} & \targ
    & \targ & \ctrl{-1} & \ctrl{-3} & \targ     & \targ
    & \rstick{\ket{0}}\qw
}}
\bigskip

In general for a single transposition, the number of multi--controlled Toffoli gates is $4$, and the number of CNOT gates is $b(I,J)$.  Note that $1\leq b(I,J)\leq n$.  So if the permutation $\pi:[[2^n]]\rightarrow[[2^n]]$ is decomposed into $L$ transpositions, the total number $T_1(n)$ of multi--controlled Toffoli and CNOT gates used satisfies
$$5L\leq T_1(n)\leq(4+n)L,$$
if we discount contractions of the type described above.  In particular, we can always find a decomposition into at most $2^n-1$ transpositions.  Whence the total number of gates is bounded above by $O(n2^n)$.

\section{Bit--wise adjacent decompositions}

We will say two integers are {\bf bit--wise adjacent} if their Hamming distance is exactly one.  And we say that a transposition $(I,J)$ is bit--wise adjacent provided that $I,J$ are bit--wise adjacent.

In the next subsection, we will give a construction that produces a bit--wise adjacent decomposition of a given transposition.  It follows that, in general, any permutation $\pi$ can be written as a product of bit--wise adjacent transpositions, which we call a {\bf bit--wise adjacent decomposition}.  We define $\|\pi\|_2$ to be smallest number of transpositions among all possible bit--wise adjacent decompositions of $\pi$.

\subsection{Bit--wise adjacent transpositions}
\begin{fact}\label{fact:bitadj}
For any transposition $(I,J)$, $\|(I,J)\|_2=2b(I,J)-1$.
\end{fact}

One possible construction is as follows.  First, find a sequence of distinct bit--wise adjacent integers
$$I^{(0)}\doteq I,I^{(1)},\dots,I^{(k)}=J,$$
where $k\doteq b(I,J)$.  Indeed, let $\iota(1)<\iota(2)<\dots<\iota(k)$ denote the indices of the non--zero bits in $I\oplus J$.  That is, $(I\oplus J)_{\iota(j)}=1$ for $1\leq j\leq k$.  Then define
$$I^{(j)}\doteq I^{(j-1)}\oplus 2^{\iota(l)},
  \quad\text{for $1\leq j\leq k$}.
$$
We then have the decomposition of transposition $(I,J)$ into a product of $2k-1$ bit--wise adjacent transpositions:
$$(I,J)=\sigma_\iota^{-1}\cdot(I^{(0)},I^{(1)})\cdot \sigma_\iota,
  \quad\text{where}\quad
  \sigma_\iota\doteq (I^{(1)},I^{(2)})\cdots(I^{(k-1)},I^{(k)}).
$$
Note that $\sigma_\iota^{-1}$ is obtained from $\sigma_\iota$ by inverting the order of transpositions.  Both are necessarily cycles.

For example, to decompose the transposition $(7,12)$ we note that $7\oplus 12=0111_2\oplus 1100_2=1011_2$. So that $\iota(1)=0$, $\iota(2)=1$, and $\iota(3)=3$.  We have the bit--wise adjacent sequence $I^{(0)}=0111_2=7$, $I^{(1)}=0110_2=6$, $I^{(2)}=0100_2=4$, $I^{(3)}=1100_2=12$.  Whence $\sigma_\iota=(I^{(1)},I^{(2)})(I^{(2)},I^{(3)})=(6,4)(4,12)$, which yields the decomposition
$$(7,12)=(4,12)(6,4)(7,6)(6,4)(4,12).$$

The above construction shows that $\|(I,J)\|_2\leq 2k-1$.  We cannot do better.  Indeed, any bit--wise adjacent decomposition of $(I,J)$ must involve at least $k-1$ letters that are neither $I$ nor $J$.  Moreover, each of these letters must occur at least twice, for a total of at least $2(k-1)$ transpositions.  On the other hand, it is well--known that every decomposition of a permutation into transpositions has the same parity: an even or odd number of transpositions.  Because the parity of $(I,J)$ is odd, there are thus at least $2(k-1)+1=2k-1$ transpositions.

\subsection{Transposition reduction}

As indicated in the previous subsection, one possible way of constructing a bit--wise adjacent decomposition of a permutation is to first decompose it into a product of transpositions and then apply Fact~\ref{fact:bitadj} to each of these transpositions.  In particular, if $\pi$ is a permutation on $2^n$ letters, we can find a decomposition with at most $2^n-1$ transpositions.  Fact~\ref{fact:bitadj} then implies the upper bound
\begin{equation}\label{eq:pi2bound}
  \|\pi\|_2\leq(2n-1)(2^n-1).
\end{equation}
In this subsection, we give a greedy strategy to reduce the number of bit--wise adjacent transpositions obtained in this way.  However, I do not know if this strategy always yields the smallest possible number of bit--wise adjacent transpositions.

For a permutation $\pi$, we abuse notation and write $s\in\pi$ if $s$ is a letter in $\pi$ for which $\pi(s)\neq s$.  Define the distance between a letter $x$ (not necessarily in $\pi$) and $\pi$ to be the largest Hamming distance between $x$ and all the letters in $\pi$:
$$b(x,\pi)\doteq\max\{b(x,s)\mid s\in\pi\}.$$
And we will call any letter $x$ that minimizes the distance to $\pi$ a {\bf bit--wise minimal letter} for $\pi$.  That is, if $b(x,\pi)\leq b(y,\pi)$ for all letters $y$.  For example, if $\pi=(1,4)(2,8)$, then using binary representations $1=0001_2$, $2=0010_2$, $4=0100_2$, $8=1000_2$ we see that $b(0,\pi)=1$.  So 0 is a bit--wise minimal letter for $\pi$ (and is in fact the unique such letter).

Because a permutation can canonically be written as a product of disjoint (hence commuting) cycles, it suffices to consider only cycles of length greater than two.  To this end, suppose that $\sigma=(s_0,\dots,s_{m-1})$ is a cycle.  First assume that $\sigma$ contains a bit--wise minimal letter, which without loss of generality we may assume is $s_0$.  In this case, we have the decomposition
$$\sigma=(s_0,s_{m-1})\cdots(s_0,s_2)(s_0,s_1).$$
Now replace each transposition with its bit--wise adjacent transposition decomposition from Fact \ref{fact:bitadj}.  We will call the resulting bit--wise adjacent decomposition the {\bf internal minimal decomposition} of $\sigma$.

On the other hand, suppose that $\sigma$ does not contain a bit--wise minimal letter.  That is if $x$ is a bit--wise minimal letter for $\sigma$, then necessarily $x\not\in\sigma$.  In this case, we have the decomposition
$$\sigma=(x,s_0)(x,s_{m-1})\cdots(x,s_1)(x,s_0).$$
Again, we appy Fact~\ref{fact:bitadj} to each of the transpositions in the decomposition.  We call the resulting decomposition the {\bf external minimal decomposition} of $\sigma$.

As an example, consider the cycle $\sigma=(0,2,5)$.  One possible decomposition into transpositions is $\sigma=(0,2)(2,5)$.  Observe that $b(0,2)=1$ and $b(2,5)=3$.  Using Fact~\ref{fact:bitadj}, we then get the bit--wise adjacent decomposition
$$\sigma=(0,2)(1,5)(3,1)(2,3)(3,1)(1,5).$$
On the other hand, $0$ is a bit--wise minimal letter for $\sigma$ with $b(0,\sigma)=2$.  The internal minimal decomposition of $\sigma$ is then
$$\sigma=(0,5)(0,2)
        =(1,5)(0,1)(1,5)(0,2).
$$
A brute force search over all possible bit--wise adjacent decompositions of $\sigma$ in fact yields $\|\sigma\|_2=4$.

In sum, our strategy for reducing the number bit--wise adjacent transpositions used to decompose a permutation $\pi$ is to decompose it as a product of disjoint cycles (natural algorithm).  And for each cycle, find its internal or external minimal decomposition (whichever applies).  However, this strategy is expensive: searching for a bit--wise minimal letter for a cycle takes $O(2^{2n})$ time in the worst case.

\section{Circuits with no ancillae}

Suppose that $(I,J)$ is a bit--wise adjacent transposition, with $I\oplus J=2^i$.  That is, the binary representations of $I$ and $J$ differ only by the bit at index $i$.  Then $(I,J)$ is represented by the multi--controlled Toffoli gate on $n$ qubits with target qubit at index $i$, and controlled by the common bits of $I$ and $J$ on the remaining qubits.  For example, $(4,6)$ is represented by the single gate quantum circuit\par
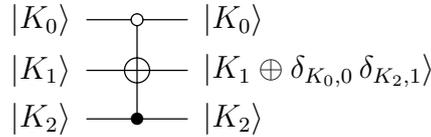
\centerline{\Qcircuit @R=12pt @C=14pt{
  \lstick{\ket{K_0}} & \ctrlo{1} & \rstick{\ket{K_0}}\qw\\
  \lstick{\ket{K_1}} & \targ  & \rstick{\ket{K_1\oplus\delta_{K_0,0}\,\delta_{K_2,1}}}\qw\\
  \lstick{\ket{K_2}} & \ctrl{-1}     & \rstick{\ket{K_2}}\qw
}}
\medskip

In general, we obtain a quantum circuit representation of a permutation by using a bit--wise adjacent decomposition.  For instance, the permutation $\pi=(0,7,12)(4,5)$ has bit--wise adjacent decomposition
$$(0,7,12)(4,5)=(0,4)(4,12)(5,7)(4,5)(5,7)(4,0)(4,5)$$
(which is in fact optimal).  Indeed, $4$ is a minimal letter for the cycle $\sigma=(0,7,12)$ with $b(4,\sigma)=2$, so that $\sigma=(4,0)(4,12)(4,7)(4,0)$ is an external minimal decomposition.  Moreover, $(4,7)=(5,7)(4,5)(5,7)$ is a bit--wise adjacent decomposition of the only remaining non bit--wise adjacent transposition.  This gives the $4$--qubit circuit for $\pi$:
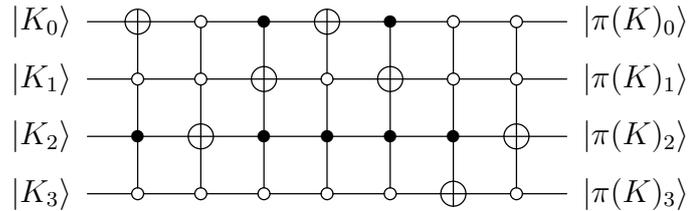
\par
\centerline{\Qcircuit @R=12pt @C=14pt{
  \lstick{\ket{K_0}}
    & \targ & \ctrlo{1} & \ctrl{1} & \targ & \ctrl{1} & \ctrlo{1} & \ctrlo{1}
    & \rstick{\ket{\pi(K)_0}}\qw\\
  \lstick{\ket{K_1}}
    & \ctrlo{-1} & \ctrlo{1} & \targ & \ctrlo{-1} & \targ & \ctrlo{1} & \ctrlo{1}
    & \rstick{\ket{\pi(K)_1}}\qw\\
  \lstick{\ket{K_2}}
    & \ctrl{-1} & \targ & \ctrl{-1} & \ctrl{-1} & \ctrl{-1} & \ctrl{1} & \targ
    & \rstick{\ket{\pi(K)_2}}\qw \\
  \lstick{\ket{K_3}} & \ctrlo{-1}
    & \ctrlo{-1} & \ctrlo{-1} & \ctrlo{-1} & \ctrlo{-1} & \targ & \ctrlo{-1}
    & \rstick{\ket{\pi(K)_3}}\qw
}}
\bigskip

By definition, we can represent any permutation $\pi$ using $\|\pi\|_2$ multi--controlled Toffoli gates.  So by equation \eqref{eq:pi2bound}, this requires $O(n2^n)$ gates, but with asymptotically twice as many gates as used in the one ancilla circuits described in section \ref{sec:oneancilla}.



\begin{thebibliography}{66}
\bibitem{Adhikari} Bibhas Adhikari, {\it Random sampling of permutations using quantum circuits,} IEEE Transactions on Computer--Aided Design of Integrated Circuits and Systems, doi: 10.1109/TCAD.2025.3591735.
\bibitem{Khandelwal} Ankit Khandelwal, Handy Kurniawan, Shraddha Aangiras, \"{O}zlem Salehi, and Adam Glos, {\it Classification and transformation of quantum circuit decompositions for permutation operations,}  Quantum Inf Process 23, 322 (2024).
\bibitem{Miller} D. Michael Miller, Robert Wille, and Zahra Sasanian, {\it Elementary Quantum Gate Realizations for Multiple--Control Toffoli Gates,} 2011 41st IEEE International Symposium on Multipe--Valued Logic, Tuusula, Finland, 2011, 288--293.
\bibitem{Shende} Vivek V. Shende, Aditya K. Prasad, Igor L. Markov, John P. Hayes, {\it Synthesis of reversible logic circuits,} IEEE Transactions on Computer--Aided Design of Integrated Circuits and Systems, vol. 22, no. 6 (2003), 710--722.

\end{thebibliography}
\end{document}